\documentclass[%
 aip,
cp,  
 amsmath,amssymb,
 reprint,%
]{revtex4-2}
\usepackage{graphicx}
\usepackage{dcolumn}
\usepackage{bm}

\usepackage[utf8]{inputenc}
\usepackage[T1]{fontenc}
\usepackage{mathptmx} 

\begin{document}

\title{First Principle Analysis of the Magnetism and Electronic Structure of Fe$_{2}$\textit{X}Si (\textit{X}=Ti, V)}
\vspace{1em} 

\author{Hanieh ~Kachooee} 
\email[Corresponding author: ]{hanieh.kachooee@gmail.com}
 \affiliation{
  \mbox{Department of Physical Science, Bergen Community College, Paramus, New Jersey 07652, USA}
}
\author{Cindy ~Kim}%
 \email{mkim159526@me.bergen.edu}
\affiliation{
  \mbox{Department of Physical Science, Bergen Community College, Paramus, New Jersey 07652, USA}
}
\author{Jason ~Carbajal}%
\email{jcarbajal152832@me.bergen.edu}
\affiliation{
  \mbox{Department of Physical Science, Bergen Community College, Paramus, New Jersey 07652, USA}
}
\author{K.~Hettiarachchilage}
\email{kalani.hettiarachchilage@csi.cuny.edu}
\affiliation{%
 \mbox{Department of Physics and Astronomy, College of Staten Island, 2800 Victory Blvd., Staten Island, New York 10314, USA}
}%
\author{N.~Haldolaarachchige \vspace{1em} }

\email{nhaldolaarachchige@bergen.edu}

\affiliation{
 \mbox{Department of Physical Science, Bergen Community College, Paramus, New Jersey 07652, USA}
}
 

\begin{abstract}
The electronic, magnetic, and mechanical properties of Fe$_{2}$\textit{X}Si, where \textit{X} is titanium (Ti) and vanadium (V), are investigated using computational methods. Volume optimization reveals that the ground state of Fe$_{2}$TiSi is a nonmagnetic narrow-band-gap semiconductor, and that of Fe$_2$VSi is a ferrimagnetic metal. The negative formation energies of both materials confirm the stability of their crystal structures. Mechanical properties confirm the static stability of the crystal structure and suggest that the titanium compound is ductile; however, the vanadium material is brittle. Density of states and band structure studies confirm the nonmagnetic semiconducting nature with an indirect band gap at $\Gamma$ and $X$ for the titanium material and the ferrimagnetic-metallic nature of the vanadium material. Electronic and magnetic properties of the materials were investigated with applied tension (negative pressure) and compression (positive pressure) to the crystal structure. The pressure study on Fe$_{2}$TiSi shows a tunable band gap and a possible semiconductor-metal transition, and Fe$_{2}$VSi shows a tunable magnetic moment and a possible low-spin/high-spin magnetic transition.   

\end{abstract}

\maketitle

\section{Introduction}
Heusler compounds are a class of intermetallic materials with diverse electronic and magnetic properties. The common form of Heusler compounds is a ternary intermetallic compound. They have a cubic structure composed of either a half-Heusler (XYZ), a system limited to one magnetic sublattice, or a full-Heusler (X$_2$YZ), which can have two magnetic sublattices\cite{sharma2024heusler}. The X and Y variables represent transition metals, and Z represents a \textit{p}-group element. Heusler compounds compose a large family with large diversity; there are thousands of known variants with different elemental compositions \cite{GRAF20111}. Due to the variety, the electronic, magnetic, and structural properties of the compound differ, which also allows a diverse range of physical properties to be observed within one of its members \cite{Elphick31122021}. Heusler compounds permit great flexibility of tuning physical properties via electronic band engineering \cite{TAVARES2023101017}. Their highly tunable crystal and electronic structures enable control of functionalities such as magnetism, semiconductivity, half-metallicity, and superconductivity \cite {GRAF20111,Elphick31122021}.

In general, Heusler compounds with magnetic elements such as Fe, Co, and Ni tend to be metallic because of high electron density near the Fermi energy and \textit{d}-electron delocalization, but some show interesting half-metallic magnetism, such as Fe$_2$YSi \cite{Luo_2007}. However, in certain Heusler systems, small changes in their composition can lead to different electronic and magnetic properties, enabling rare combinations of semiconducting and ferromagnetic behavior \cite{gurbuz2023spinpolarizedtwodimensionalelectronholegas,article_1654241}. Full-Heusler compounds are particularly well suited for this purpose. Unlike half-Heuslers, which contain a vacant lattice site and often exhibit weaker magnetism, full-Heuslers have all lattice sites fully occupied. This occupancy allows stronger magnetic interactions, enabling ferromagnetic alignment and higher total magnetic moments \cite{sharma2024heusler}.
For instance, the full-Heusler compound Fe$_2$CrAl can maintain ferromagnetic ordering of its Fe and Cr atoms while preserving a semiconducting gap \cite{DAHMANE2016167,10.1063/1.4710435,10.1063/5.0016904}. Chemical substitution can tune the balance between conduction and magnetism, enabling access to electronic states that are normally incompatible in conventional semiconductors. By adjusting electron count, band structure, and magnetic interactions, researchers can design materials with tailored electronic and magnetic properties.
Because semiconducting ferromagnets are rare, they attract interest for potential applications in spintronics and thermoelectrics \cite{10.1063/1.4802504,PhysRevB.103.054407}. 
In particular, Fe-based full-Heusler compounds are promising due to their composition of abundant, low-cost elements and high thermal stability. For example, Fe$_2$TaAl and Fe$_2$TaGa exhibit semiconducting band gaps while satisfying mechanical stability criteria, indicating both electronic and mechanical functionality \cite{article}. These findings motivate further investigation into Fe-based Heusler compounds as candidates for multifunctional materials that combine semiconducting, magnetic, and thermoelectric properties within a single, tunable crystal framework.

Despite studies on Fe-based Heuslers, the ternary compounds Fe$_2$TiSi and Fe$_2$VSi remain underexplored \cite{guo2018, Jong2016, Jong2017}. This system contains \textit{d}-group elements, which introduce moderate electron correlations and hybridization effects. A Fe$_2$TiSi pure bulk phase has not yet been reported experimentally, and theoretical studies of the ground state of Fe$_2$\textit{X}Si, where \textit{X}=Ti and V, seem to lack the use of different exchange-correlation functionals \cite{abuova2022,Liang2025,Meinert2014}. Therefore, here we provide a comparative study of Fe$_2$TiSi and Fe$_2$VSi full-Heusler compounds. In this study, we perform a density functional theory (DFT) study of Fe$_2$\textit{X}Si where \textit{X}=Ti and V. Using volume optimization and self-consistent field calculations, we investigate their ground-state properties, electronic band structures, density of states, magnetic behavior, and mechanical properties. Then the effect of pressure on the electronic structure is investigated. 

\section{Computational Method}
All simulations were performed using the Expanse and Bridges high-performance computer clusters \cite{SDSC_Expanse_UserGuide_2025,access}. Quantum ESPRESSO (QE) was utilized to analyze the Fe-(Ti/V)-Si systems, and the Material Project and Materials Cloud platforms were used to extract structural data and potential functions \cite{Giannozzi_2009,Talirz_2020,Jain2013}. Volume optimization was done by using the variable-cell relaxation (vc-relax) method with QE. Three spin configurations, nonmagnetic (NM, no spin orientations), ferromagnetic (FM, atomic spins align in parallel), and antiferromagnetic (AFM, nearest-neighbor atomic spins align in antiparallel directions), were taken into account for each compound. The convergence was performed with the plane-wave cutoff of 60 Ry for kinetic energy and 600 Ry for the charge density. Additionally, the self-consistent-field (SCF) convergence threshold was set at $10^{-9}$ Ry. The lowest total energy of the three spin configurations (NM, FM, AFM) was used to determine in which case the compound was at the ground state. For the magnetic ground state, the Hubbard potential was applied within the generalized gradient approximation (GGA+U). Hubbard potential U values are set to U=2.00 eV for Fe atoms and U=1.50 eV for V atoms. These U values are in good agreement with the literature for transition-metal compounds with moderate electron correlations \cite{prb22}. If the ground state showed a band gap, then the modified Becke-Johnson (mBJ) method was used to validate the band gap. The $9 \times9 \times 9$ k-point matrix was used. To get the density of states (DOS), a non-self-consistent field (NSCF) calculation was performed, followed by a DOS calculation. Formation energy was calculated using the ground-state energies of the compounds and constituent elements. Mechanical properties were calculated with the ElaStic software package\cite{elastic}, which cooperated with QE.

\section{Results and Discussion}
\subsection{Ground State with Relaxed Volume Optimization}

 \begin{figure}[!htbp]
    \centering
    \includegraphics[width=0.6\linewidth]{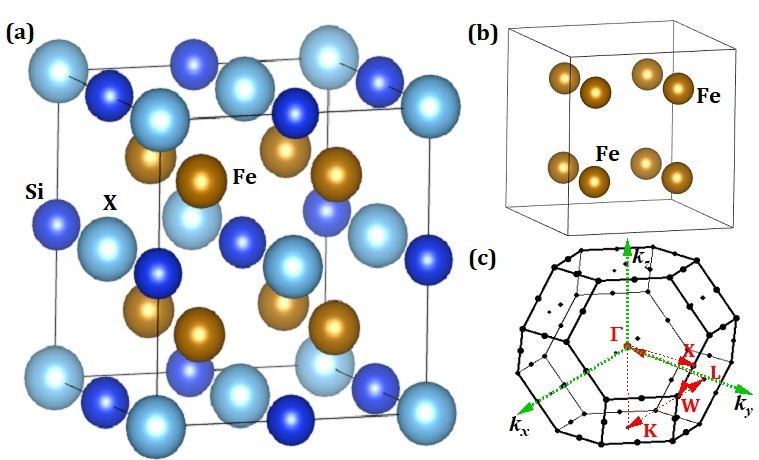}
    \caption{(a) Cubic full-Heusler structure of Fe$_2$\textit{X}Si (\textit{X}=Ti, and V) showing Si (dark blue), Ti and V (light blue), and Fe atoms (bronze). (b) Fe atoms in a cubic sublattice. (c) The first Brillouin zone and high-symmetry points. The green arrows represent reciprocal lattice axes.}
    \label{fig:Fig1}
\end{figure}

 Figure \ref{fig:Fig1}(a) shows the standard cubic unit-cell structure of full-Heusler compounds Fe$_2$\textit{X}Si, which crystallizes into the L2$_1$ structure (space group of 225, Fm$\bar{3}$m). Ti and V atoms (light blue) are positioned at (0, 0, 0), and Si atoms (dark blue) are positioned at (0.5, 0.5, 0.5). The arrangement of \textit{X} and Si atoms forms a rock-salt-like sublattice. The Fe atoms (bronze) take the tetrahedral voids at (0.25, 0.25, 0.25) and (0.75, 0.75, 0.75). Figure \ref{fig:Fig1}(b) shows the Fe atom cubic sublattice inside the full-Heusler cubic unit cell. Figure \ref{fig:Fig1}(c) shows the first Brillouin zone in the reciprocal space of the face-centered cubic (FCC) lattice. The red dotted arrows in Fig. \ref{fig:Fig1}(c) represent the high-symmetry path $\Gamma$ $\rightarrow$ X $\rightarrow$ W $\rightarrow$ K $\rightarrow$ $\Gamma$ $\rightarrow$ L that is used for the electronic band structure calculations. The selection of a high-symmetry path plays an important role when identifying the nature of the band gap, whether direct or indirect, of a semiconducting compound, and generally novel electronic properties.  

\begin{table}[htbp]
\centering
\caption{Volume optimization results for Fe$_2$TiSi and VFe$_2$Si using VC-relax.}
\label{tab:volume_optimization}
\small
\begin{tabular}{lcccc}
\hline\hline
Method & $a$ (\AA) & $V$ (\AA$^3$) & $E_0$ (eV) & $\Delta E$ (eV) \\
\hline
\multicolumn{5}{l}{\textbf{Fe$_2$TiSi}} \\
NM  & 5.688 & 184.05 & $-48389.657$ & 0.000 \\
FM  & 5.691 & 184.32 & $-48389.634$ & 0.023 \\
AFM & 5.688 & 184.06 & $-48389.634$ & 0.023 \\
\hline
\multicolumn{5}{l}{\textbf{Fe$_2$VSi}} \\
NM  & 5.652 & 180.54 & $-49976.480$ & 16.100 \\
FM  & 5.659 & 181.28 & $-49992.580$ & 0.000 \\
AFM & 5.887 & 203.98 & $-49989.294$ & 3.286 \\
\hline\hline
\end{tabular}
\end{table}

Table \ref{tab:volume_optimization} shows a summary of volume optimization to investigate the ground state of Fe$_2$\textit{X}Si. Three spin configurations, NM: nonmagnetic (no spin alignments), FM: ferromagnetic (all spins in parallel), and AFM: antiferromagnetic (nearest spins align in antiparallel), are used to identify the ground state. The table shows the lattice parameter, unit-cell volume, total energy (E$_0$), and energy difference (i.e., $\Delta$E between the lowest energy state and other states). When total energy is compared between different configurations, it is evident that the ground state of the Fe$_2$TiSi is nonmagnetic and Fe$_{2}$VSi is ferrimagnetic (antiparallel spin alignment between atoms with a net positive magnetic moment). The ground-state lattice constants of Fe$_2$TiSi and Fe$_2$VSi are in good agreement with previously reported values in the literature \cite{Jong2016,abuova2022}. 

These ground states can be investigated within the valence electron counting (VEC) based Slater-Pauling (SP) rule that gives the total magnetic moment of a full-Heusler, X$_2$YZ, where X and Y are \textit{d} group elements, and Z is a main group element:
\begin{equation}
    M_t = Z_t - 24.
    \label{eq:slater_pauling}
\end{equation}
where M$_t$ is the total magnetic moment per formula unit and Z$_t$ is the total valence electrons per (f.u.).   
According to the SP rule the Fe$_2$TiSi compound should have zero total magnetic moment because the total VEC is Z$_t$=24. This agrees well with our calculations. VEC for the Ti compound is as follows: four electrons from Ti ($3d^2 4s^2$), sixteen from the two Fe atoms ($2 \times 3d^6 4s^2$), and four from Si ($3s^2 3p^2$). 

\begin{table}[ht]
    \centering
    \caption{Magnetic moments of the Fe$_2$VSi system.}
    \label{tab:mm}
    \begin{tabular}{|l|c|c|c|c|}
        \hline
        & \multicolumn{4}{|c|}{Magnetic moment ($\mu_B$/f.u.)} \\
        \hline
        Method & Total & Fe & V & Si \\
        \hline
        GGA   & 0.9301 & 0.5238 & -0.0978 & -0.0197 \\
        \hline
        GGA+U & 0.9450 & 1.0508 & -0.8271 & -0.0528 \\
        \hline
    \end{tabular}
\end{table}

Table \ref{tab:mm} shows calculated magnetic moment results of Fe$_2$VSi with GGA and GGA+U.
The SP rule predicts that the vanadium compound should have a magnetic ground state with M$_t$=1.0 $\mu_B$ because the VEC of Fe$_2$VSi is Z$_t$=25. The VEC of the vanadium compound is as follows: five electrons from V ($3d^3 4s^2$), sixteen from the two Fe atoms ($2 \times 3d^6 4s^2$), and four from $Si$ ($3s^2 3p^2$). Our calculation shows that GGA results in a total magnetic moment of 0.93 $\mu_B$ per f.u., which is close to the predicted value of the SP rule and recent literature \cite{abuova2022}. However, the GGA method underestimates electron correlations and therefore results in a lower magnetic moment on correlated electron systems \cite{cohen2012,jyoti_2023}. This is particularly evident with the magnetic moment per Fe atom (0.5238 $\mu_B$ per f.u.), which is a very low value for the localized magnetic moment per Fe atom. 
Our GGA+U calculation results in a slightly higher total magnetic moment (1.05 $\mu_B$ per f.u.), which is in great agreement with the SP rule and with reported values in literature \cite{abuova2022}. Additionally, the GGA+U method resulted in higher magnetic moments per atom for the Fe$_{2}$VSi compound, with a magnetic moment per Fe atom M(Fe/f.u.)=1.0508 $\mu_B$, a magnetic moment per V atom M(V/f.u.)=-0.8271 $\mu_B$, and a magnetic moment per Si atom M(Si/f.u.)=-0.0528 $\mu_B$. This leads to a total magnetic moment M(total/f.u.)=0.9450 $\mu_B$. There are not reported magnetic moments with GGA+U in the literature for this material. The GGA+U results from this study agree well with the experimentally measured total magnetic moment of this compound and other similar full-Heusler materials, such as Co$_2$RuSi ~\cite{abuova2022, 2017JM}. The higher total magnetic moment with the Hubbard potential (U) confirms the effect of moderate electron correlations. Generally, the GGA method is not able to localize \textit{d}-electrons in V and Fe atoms, which leads to a reduced or quenched total magnetic moment. When the Hubbard potential U is applied (GGA+U), it incorporates localized \textit{d}-electrons in V and Fe atoms, which leads to a higher total magnetic moment.     
When comparing the sign of the magnetic moment of each atom, the Fe moment is positive and the V and Si moments are negative. This suggests that V and Si spins are antiparallel to the Fe spin. Fe$_{2}$VSi has a positive total magnetic moment per formula unit, which means it should be ferrimagnetic (FiM, antiparallel spin between Fe and V). However, the ground state has all Fe spins aligned in the same direction (parallel), which means that Fe atoms have ferromagnetic (FM) spin coupling within the Fe atom sublattice. 

\subsection{Formation Energy}
Formation energy is calculated by using the total energy of the compounds Fe$_2$\textit{X}Si (\textit{X}=Ti and V) and the total energies of each element, Ti, V, Fe, and Si. Formation energy is an important parameter to investigate the thermodynamic stability of the crystal structure. Generally, the formation energy of a compound phase must be negative at thermal equilibrium with respect to the elemental phases that constitute the material. Formation energy was calculated using the following equation,
\begin{equation}
\begin{aligned}
E_{f} &= \frac{E_{Fe_2XSi} - N_XE_X - N_{Fe}E_{Fe} - N_{Si}E_{Si}}{N} \\
&= \frac{E_{Fe_2XSi} - E_X - 2E_{Fe} - E_{Si}}{4},
\end{aligned}
\label{eq:fe}
\end{equation}
where $E_f$ is the formation energy, $E_{Fe_2\textit{X}Si}$ is the total energy of the full Huesler compounds (Fe$_2$TiSi and Fe$_2$VSi), and the energy of elemental phases is given by $E_X$ ( $X$= Ti or V), $E_{Fe}$, and $E_{Si}$. $N_X=1$ represents one Ti or V atom per f.u., $N_{Fe}=2$ represents two Fe atoms per f.u., and $N_{Si}=1$ represents one Si atom per f.u.

Total energy calculations were done with an optimized cubic unit cell for Fe$_2$\textit{Ti}Si and Fe$_2$VSi Heusler compounds with space group 225 (Fm$\bar{3}$m). Total energy calculations of elemental phases were done with the unit cells of vanadium with space group 229 (Im$\bar{3}$m), titanium with space group 194 (P6$_3$/mmc), iron with space group 229 (Im$\bar{3}$m), and silicon with space group 227 (Fd$\bar{3}$m). Formation energy was calculated by using Eq. (\ref{eq:fe}) and resulted in $E_f$(Fe$_2$TiSi)$=-1.04$ eV/atom and $E_f$(Fe$_2$VSi)$=-1.01$ eV/atom. These negative formation energies for both compounds suggest that these full-Heusler compounds are stable at thermal equilibrium and fall within the acceptable values reported in the literature \cite{jack, jack2}. 
\subsection{Mechanical Properties and Static Stability}
Table \ref{tab:elastic_properties} shows the analysis and comparison of elastic and mechanical properties of full-Heusler Fe$_2$\textit{X}Si (\textit{X}=Ti and V). Both compounds show mechanical stability as they obey the following criteria: ($C_{11}-C_{12}>0$, $C_{11}-2C_{12}>0$, and $C_{44}>0$), which confirms the stable structures \cite{jack,laptsb}. Our calculation shows that Fe$_2$TiSi is significantly stiffer when compared with Fe$_2$VSi because Young's modulus and the shear modulus of Fe$_2$TiSi are higher than those of Fe$_2$VSi. The materials' ductility vs. brittleness can be checked with the bulk-to-shear modulus ratio (B/G) according to Pugh's criterion (ductile when B/G $>$ 1.75 and brittle when B/G $<$ 1.75). Our calculation suggests that Fe$_2$TiSi is a brittle (B/G = 1.539) compound and Fe$_2$VSi is a ductile (B/G = 3.786) compound. 
\begin{table*}[ht]
\centering
\caption{Comparison of elastic and mechanical properties of Fe$_2$TiSi and Fe$_2$VSi.}
\label{tab:elastic_properties}
\begin{tabular}{lccc}
\hline\hline
\textbf{Property} & \textbf{Fe$_2$TiSi} & \textbf{Fe$_2$VSi} & \textbf{Unit} \\
\hline
\multicolumn{4}{c}{\textit{Elastic Constants}} \\
$C_{11}$ & 347.851 & 301.130 & GPa \\
$C_{12}$ & 106.835 & 139.555 & GPa \\
$C_{44}$ & 122.420 & 37.259 & GPa \\
\hline
\multicolumn{4}{c}{\textit{Mechanical Stability Criteria}} \\
$C_{11} - C_{12}$ & 241.016 & 161.575 & GPa \\
$C_{11} + 2C_{12}$ & 561.521 & 580.240 & GPa \\
$C_{44}$ & 122.420 & 37.259 & GPa \\
Mechanically stable? & Yes & Yes & -- \\
\hline
\multicolumn{4}{c}{\textit{Derived Elastic Moduli}} \\
Bulk modulus ($B$) & 187.174 & 193.414 & GPa \\
Shear modulus ($G$) & 121.652 & 51.083 & GPa \\
Young's modulus ($E$) & 299.968 & 140.849 & GPa \\
Poisson's ratio ($\nu$) & 0.233 & 0.379 & -- \\
\hline
\multicolumn{4}{c}{\textit{Mechanical Behavior}} \\
$B/G$ ratio & 1.539 & 3.786 & -- \\
Nature & Brittle & Ductile & -- \\
Cauchy pressure ($C_{12} - C_{44}$) & -15.585 & 102.296 & GPa \\
\hline
\multicolumn{4}{c}{\textit{Elastic Anisotropy}} \\
Zener ratio ($A_Z$) & 1.016 & 0.461 & -- \\
Universal anisotropy ($A^U$) & 0.000 & 0.755 & -- \\
Character & Isotropic & Anisotropic & -- \\
\hline
\multicolumn{4}{c}{\textit{Other Properties}} \\
Debye temperature ($\theta_D$) & 625.4 & 408.4 & K \\
Vickers hardness ($H_v$) & 17.0 & 1.2 & GPa \\
Density ($\rho$) & 6.39 & 6.60 & g/cm$^3$ \\
\hline\hline
\end{tabular}
\end{table*}

\vspace{1cm}

Cauchy pressure ($C_{12}-C_{44}$) is a parameter that can be used to understand the bonding character, metallic or covalent, of a compound \cite{dujana}. The titanium compound has negative Cauchy pressure, which suggests covalent bonding character, resulting in stronger \textit{p}-\textit{d} hybridization between silicon \textit{p}-orbitals and transition-metal \textit{d}-orbitals. The vanadium compound shows a metallic bonding nature with positive Cauchy pressure, which agrees with the ductile nature of this compound. 
The results suggest that the titanium compound is isotropic with a Zener ratio (A$_Z$=1.016) close to 1, while the vanadium compound shows anisotropic behavior with $A_Z<1$. The titanium compound seems to be much harder, with a Vickers hardness (H$_v$) of 17.0 GPa, compared to the vanadium compound with H$_v$=1.2 GPa. Estimated Debye temperatures ($\theta_D$) suggest stronger interatomic bonding and higher thermal conductivity in the titanium compound ($\theta_D$=625.4 K) compared to the vanadium compound ($\theta_D$=408.4 K) \cite{paula}.

\subsection{Electronic Structure}

\begin{figure}[t!]
    \centering
    \includegraphics[width=0.57\linewidth]{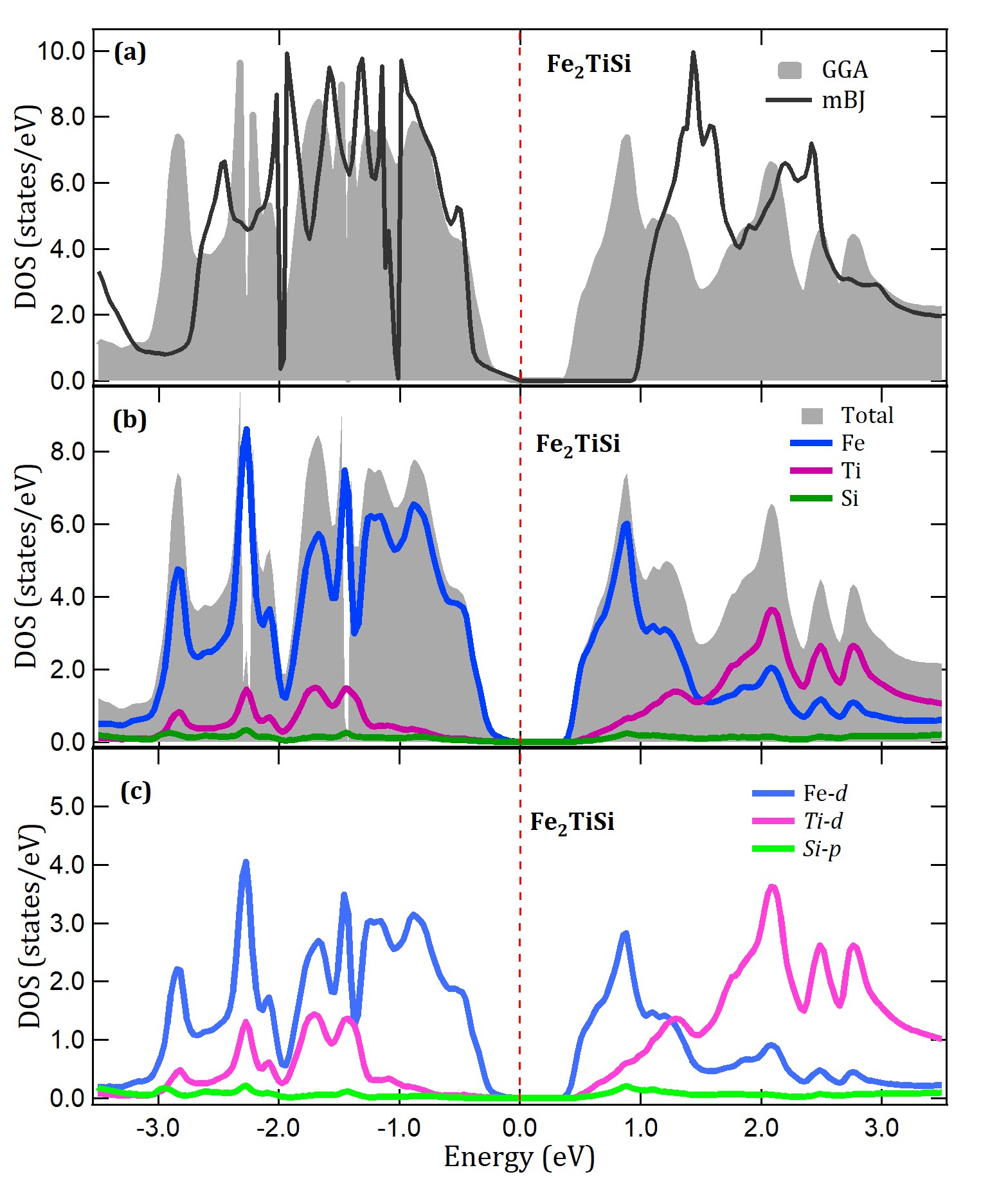}
    \caption{Density of states (DOS) and partial density of states (PDOS) of Fe$_2$TiSi. (a) Total DOS with GGA (gray shaded area) and mBJ (solid black line), (b) total (gray area) and atom projected DOS (Fe in blue, Ti in purple, Si in green), and (c) orbital projected DOS of each atom (Fe in blue, Ti in purple, Si in green).}
    \label{fig:Fig2}
\end{figure}

Figure \ref{fig:Fig2} shows the analysis of the density of states (DOS) of Fe$_{2}$TiSi. The Fermi level is set to E=0 eV on all three panels. Figure \ref{fig:Fig2}(a) shows the total DOS determined by two computational methods, GGA and MBJ. Both methods confirm that the titanium compound is a semiconductor. The estimated band gap is 0.368 eV with GGA and  0.980 eV with mBJ. The GGA result is in good agreement with the reported band gap in the literature~\cite{Jong2016}. However, in general, the GGA method underestimates the band gap of semiconductors and insulators. It has been shown that the mBJ method generally makes a good estimation of semiconductor band gaps~\cite{Tran2009}. Therefore, it can be assumed that if a pure phase of bulk TiFe$_s$Si is made experimentally, it will be a semiconductor with a band gap between 0.4 and 1.0 eV. There is no pure bulk phase experimentally reported on this system. Recently Liang et al. reported this system, but with a considerable amount of two impurity phases \cite{Liang2025}. Those impurity phases are known to be metallic, which may be the dominant reason their bulk sample is metallic. 

Figure \ref{fig:Fig2}(b) shows the projected density of states (PDOS) of each atom, and panel (c) shows the orbit projected DOS of Fe-\textit{d}, Ti-\textit{d}, and Si-\textit{p} of Fe$_2$TiSi. The PDOS of Fe dominates near the Fermi energy, which is below the valence bands, and above in the conduction bands, and panel (c) shows that it is Fe-\textit{d} orbitals. This strongly suggests that the electronic properties of this compound are dominated by hybridized Fe-\textit{d} orbitals. The titanium PDOS contribution is more significant at higher energies (>1.5 eV) in the conduction bands. The PDOS of Fe and Ti suggests that the electronic properties of this compound are dominated by the hybridization of Fe-\textit{d} and Ti-\textit{d} orbitals. The silicon PDOS contribution is insignificant, and that suggests the electrons from Si-\textit{p} orbitals are deeper in the valence band and/or in strong covalent bonds. 

\begin{figure}
    \centering
    \includegraphics[width=0.6\linewidth]{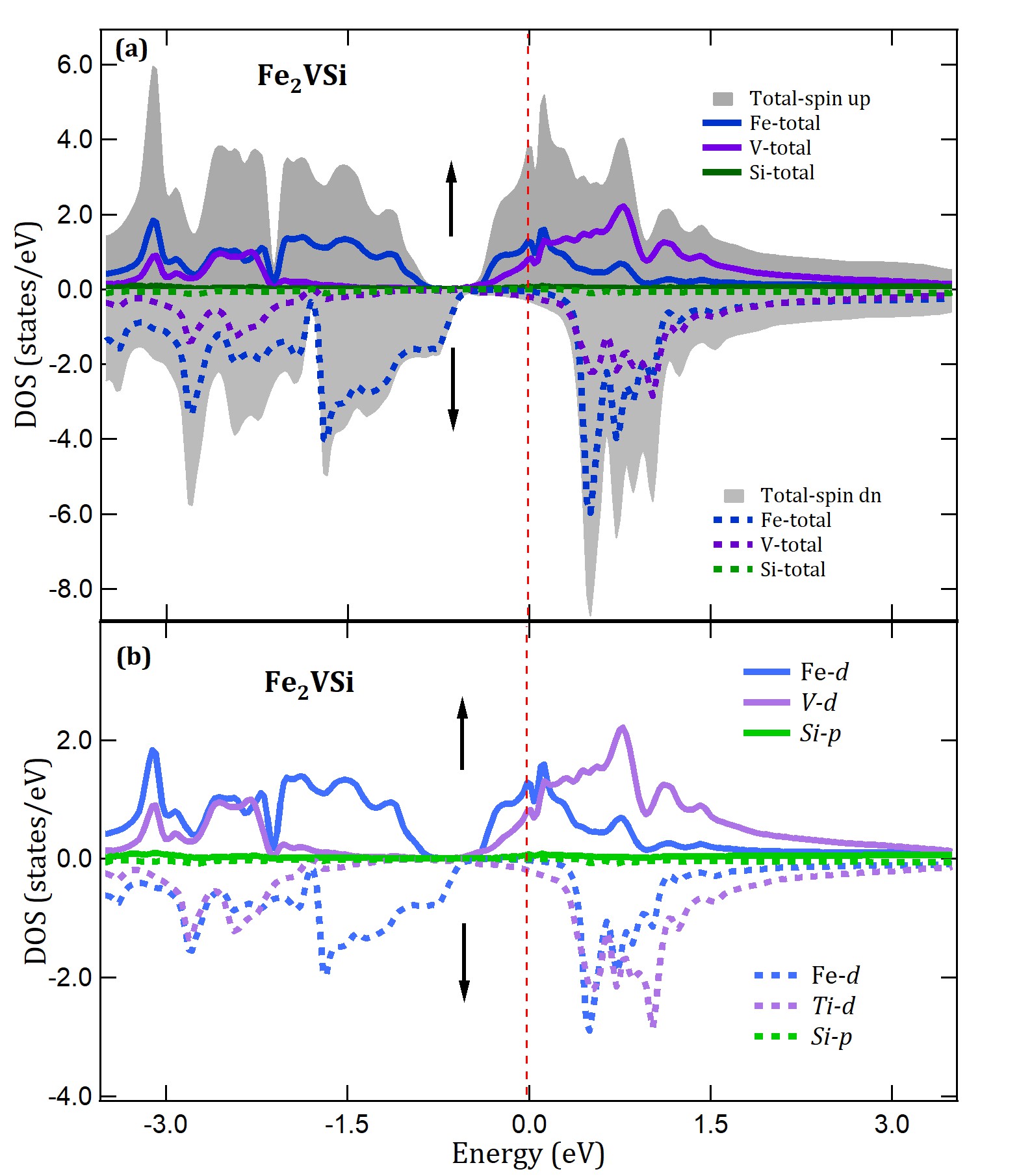}
    \caption{Spin-resolved density of states (DOS) and partial density of states (PDOS) of Fe$_2$VSi. (a) Spin-resolved total DOS (gray shaded area) and spin-resolved atom projected total DOS (Fe is in blue, V in purple, and Si in green). Spin up is portrayed as positive, spin down as negative. (b) The dominant PDOS of each atom.}
    \label{fig:Fig3}
\end{figure}

Figure \ref{fig:Fig3} shows the spin-resolved total density of states (DOS) and projected density of states (PDOS) of the full-Heusler compound Fe$_{2}$VSi. In panel (a), gray indicates the total DOS of spin up (positive y-axis) and spin down (negative y-axis). This asymmetry suggests the compound should be ferromagnetic or ferrimagnetic. A detailed investigation of the magnetic moment of each atom confirmed that it is more likely to be in a ferrimagnetic ground state (see Table \ref{tab:mm}). The large DOS on the Fermi level with no band gap confirms a metallic nature. The projected total DOS of each atom is shown in panel (a), and the orbit-projected DOS of each atom is shown in panel (b). The metallic magnetic nature of this compound agrees with the Slater-Pauling (SP) rule prediction as given in Eq. (\ref{eq:slater_pauling}). The total VEC of this compound is 25, and the SP rule suggests a magnetic ground state with a total magnetic moment of 1.0$\mu_B$. Calculated results with GGA+U agree well with the SP rule and recent literature~\cite{abuova2022}. This can be attributed to the dominant PDOS of Fe-\textit{d} and V-\textit{d} near the Fermi energy, which also suggests strong hybridization of Fe-\textit{d} and V-\textit{d}.

\begin{figure}[t!]
    \centering
    \includegraphics[width=0.6\linewidth]{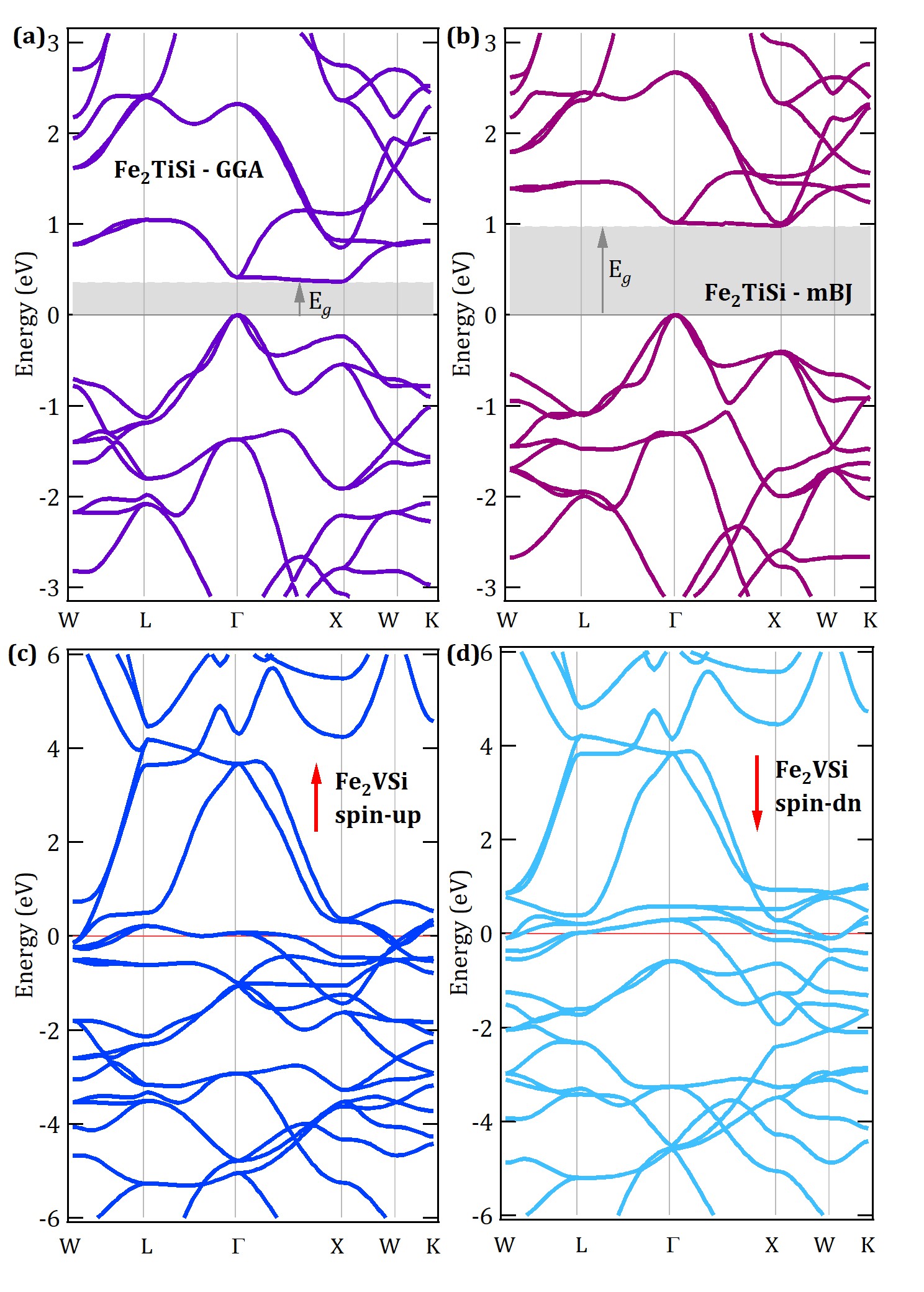}
    \caption{Electron band structures of FeX$_2$Si (\textit{X}=Ti, V). (a) Band structure with GGA and (b) with mBJ of Fe$_2$TiSi. Band structure for spin up (c) and (d) spin down of Fe$_2$VSi.}
    \label{fig:Fig4}
\end{figure}

Figure \ref{fig:Fig4} shows the bands structure of Fe$_2$\textit{X}Si (\textit{X}=Ti or V). The band structure of semiconducting Fe$_2$TiSi with the GGA method is shown in panel (a) and with the mBJ method shown in panel (b). The band gap, represented by gray, is visible in both panels (a) and (b). The band gap with the mBJ method is about two times larger compared with the band gap with GGA. As discussed earlier, the mBJ method does a better job estimating the band gap of semiconducting compounds. It appears that this material has an indirect band gap between $\Gamma$ and X, which is very close to the direct band gap at the $\Gamma$ point. Electron bands near the Fermi level are dominated by the hybridized Fe-\textit{d} orbitals. 

The spin-resolved electron band structure of Fe$_2$VSi is shown in panel (c) spin up and panel (d) spin down. The band crossing through the Fermi energy confirms the metallic behavior on both spin channels. The metallic nature is also observed with the density of states in Fig. \ref{fig:Fig3}. Significantly different band dispersion on the spin-up and spin-down channels confirms the existence of a total magnetic moment. 
Both spin channels have flat bands on or near the Fermi energy, which indicates localized magnetization. Electron bands at and near the Fermi energy are dominated by hybridized Fe-\textit{d} orbitals.
Compared to a nonmagnetic titanium system, the vanadium compound carries an extra valence electron per formula unit, which drastically changes the behavior of the two systems. 
The titanium system has 24 valence electrons, and it shows a nonmagnetic nature with a band gap. The vanadium system has 25 valence electrons, which changes the system into a ferrimagnetic metal. 

\subsection{Pressure Effect on the Band Gap of Fe$_2$TiSi}
\begin{figure}[t!]
    \centering
    \includegraphics[width=0.6\linewidth]{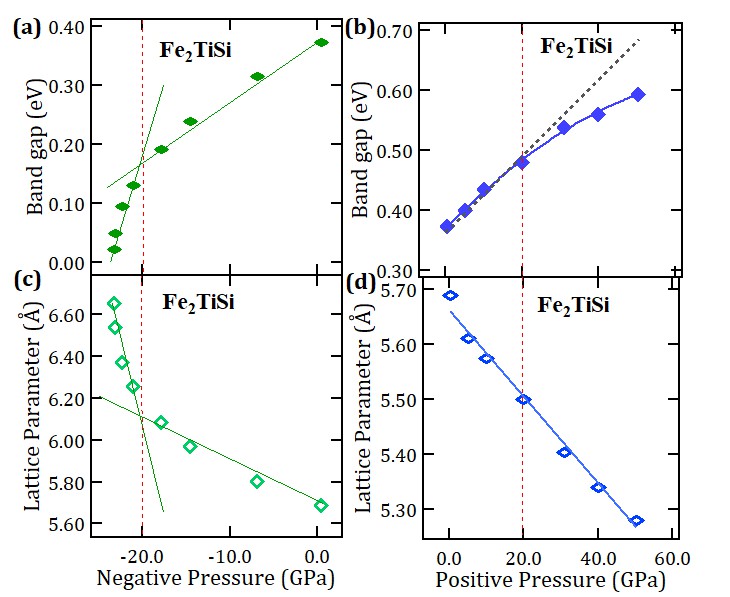}
    \caption{Pressure effect on the band gap and lattice parameter of Fe$_2$TiSi. (a) The band gap and (c) lattice parameter as a function of negative pressure, and (b) band gap and (d) lattice parameter as a function of positive pressure.}
    \label{fig:Fig5}
\end{figure}

Figure \ref{fig:Fig5} shows the band gap and lattice parameter as a function of negative and positive pressure of Fe$_2$TiSi. 
Electronic band-gap variation with respect to tension (negative pressure) is demonstrated in panel (a). As the negative pressure is applied, the band gap decreases linearly, but there is an inflection point at about -20.0 GPa. A similar feature can be observed in the lattice parameter variation in panel (c). 
The rate of change of the band gap as a function of tension (or lattice expansion) $\frac{dE_g}{dP}$ shows a significant change near the pressure $P_0$=-20 GPa. Below $P_0$, the band gap decreases rapidly, and near the pressure of -23 GPa, the band gap is 0.05 eV. This is evidence that the band gap is near the closing ($E_g$=0) below a pressure of -23 GPa. This provides strong evidence of a possible semiconductor-metal phase transition. 
As the unit cell expands with applied tension (negative pressure), the orbital overlap (or hybridization) between Fe-\textit{d}, Ti-\textit{d}, and Si-\textit{p} starts to reduce, which causes narrower bandwidths and reduced bonding-antibonding splitting. That results in the shrinking of the semiconducting band gap ($E_g$) and, eventually, at very high negative pressure ($P > - 23$ GPa) the band gap will close, which agrees well with similar pressure dependence reported in the literature \cite{galanakis}. 
On the other hand, our calculation suggests that there is no sign of a possible structural phase transition because the total energy vs. lattice parameter (and pressure) does not show an inflection point. 
Therefore, we speculate that the inflection point near -20 GPa could be some sort of electronic topological transition, such as a Lifshitz-type transition under pressure \cite{prb2013}.
 
The lattice parameter variation with positive pressure (compression) is shown in panel (d). The lattice parameter decreases with increasing compression all the way up to +50.0 GPa. Panel (b) shows the band-gap behavior within the same positive pressure range. The band gap shows a consistent increase with compression up to +50.0 GPa. There seems to be a very slight rate change of $\frac{dE_g}{dP}$ at about +20.0 GPa, but it is not significant when compared with the effect observed in panel (a) at -20.0 GPa. Therefore, it can be assumed that there is no electronic or structural phase change in the positive pressure range up to +50.0 GPa. 
The increasing band gap ($E_g$) with positive pressure is a result of an increasing orbital hybridization between transition-metal (Fe and Ti) \textit{d}-orbitals and Si-\textit{p} orbitals. That widens the gap between the conduction band minimum and valence band maximum by pushing conduction bands upward or valence bands downward. Therefore, semiconducting properties can be enhanced with applied pressure and fine-tuned for certain applications \cite{galanakis,bhat}.  

\subsection{Pressure Effect on the Magnetic Moment of Fe$_{2}$VSi}
\begin{figure}[t!]
    \centering
    \includegraphics[width=0.6\linewidth]{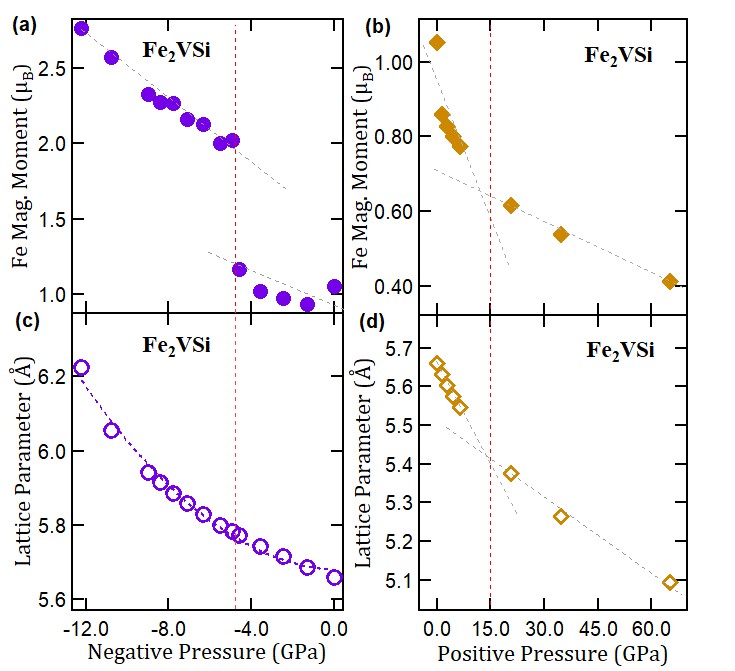}
    \caption{Pressure effect on the magnetic moment and lattice parameter of Fe$_2$VSi. The magnetic moment of the Fe atom per unit cell and the lattice parameter are shown as a function of negative pressure in (a) and (c) and positive pressure in (b) and (d).}
    \label{fig:Fig6}
\end{figure}

Figures \ref{fig:Fig6}(a) and \ref{fig:Fig6}(c) show the magnetic moment and lattice parameter as a function of negative pressure (lattice relaxation or tension). The magnetic moment increases as a function of unit-cell expansion (increasing negative pressure) from 0.0 GPa to -5.0 GPa and then -6.0 GPa to -12.0 GPa. There is an abrupt change in behavior between -4.00 GPa and -6.00 GPa pressure, which is about a 2.0$\%$ and 5.0$\%$ unit-cell volume increase relative to the volume at ambient pressure. However, the lattice parameter smoothly increases with increasing negative pressure (tension). There is no sudden change in lattice parameter as a function of negative pressure. This confirms proper structural relaxation, and there is no structural instability. 
When the unit-cell volume increases with negative pressure (tension), the atomic distance between Fe-V and Fe-Fe increases, which leads to decreasing \textit{d}-\textit{d} hybridization and decreasing bandwidth. This can cause an increasing density of states at the Fermi level (E$_F$) that leads to increasing electron localization. As a result, the Fe magnetic moment per formula unit increases with increasing negative pressure. 
It seems that negative pressure or lattice expansion transfers the system from a low-spin state to a high-spin state. The magnetic moment is below 1.25 $\mu_B$ up to about -5.0 GPa pressure or about a 2.0$\%$ increase of unit-cell volume. This seems to be a low-spin state. Then the magnetic moment sharply increases and stays above 2.0$\mu_B$ for negative pressure above -6.0 GPa as the unit-cell volume increases by more than 5$\%$. This seems to be the high-spin state. 
This indicates a pressure-driven magnetic phase transition from a low-spin state to a high-spin state between -2.0 GPa and -5.0 GPa\cite{xiao}. Further investigation of the atomic spin alignment suggests that the compound remains in a ferrimagnetic state, with the Fe spin antiparallel to the V spin. Therefore, we suggest that this magnetic transition possibly changes the strength of the Fe-V coupling and affects the correlation between the magnetic moment and the local bonding strength and hybridization \cite{abuova2022}.

Figures \ref{fig:Fig6}(b) and \ref{fig:Fig6}(d) show the behavior of the lattice parameter and magnetic moment of Fe per f.u., respectively, as a function of positive pressure (lattice compression). Both the lattice parameter and Fe magnetic moment decrease with increasing pressure. When positive pressure is applied, the unit-cell volume decreases with decreasing lattice parameter. This leads to a decrease in the separation between Fe-Fe atoms and Fe-V atoms. This means that Fe-\textit{d} and V-\textit{d} orbital overlap becomes stronger, which increases bandwidth and reduces exchange splitting. Therefore localization of the Fe-\textit{d} electrons decreases, leading to a reduction of the Fe magnetic moment per formula unit. 
The magnetic moment and lattice parameter variation with positive pressure show similar behavior with no sign of abrupt changes. 
The rate of reduction of magnetic moment and lattice parameter (unit-cell volume) per applied positive pressure changes near 15.0 GPa, nearly a 2$\%$ reduction of lattice parameter. However, this slope change is not strong because the magnetic moment values are between 1.0 $\mu_B$ and 0.4 $\mu_B$ over the whole range of applied pressure up to +50.0 GPa, which confirms the system is in the low-spin state. 

\section{Conclusion}
In summary, these intermetallic systems (Fe$_2$TiSi and Fe$_2$VSi) prove to be significant compounds due to their electronic and magnetic properties. The ground state of Fe$_{2}$TiSi proved to be a nonmagnetic semiconductor, and Fe$_{2}$VSi exhibited ferrimagnetic metallic behavior. The formation energies for both compounds suggest that these full-Heusler compounds are stable at thermal equilibrium. Both compounds met the criteria for mechanical stability, with Fe$_{2}$TiSi being significantly stiffer and having stronger \textit{p-d} hybridization between silicon \textit{p}-orbitals and transition-metal \textit{d}-orbitals. The calculated Debye temperatures suggest that the titanium compound has stronger interatomic bonding and higher thermal conductivity when compared to Fe$_{2}$VSi. Our findings show that the electronic properties of this system are dominated by the strong hybridization of Fe-\textit{d} and \textit{X-d} (X = Ti, V) orbitals near the Fermi level. The effect of physical pressure on Fe$_{2}$TiSi indicates a possible semiconductor-metal phase transition below a pressure of -23.0 GPa. A pressure study on Fe$_{2}$VSi demonstrates a possible magnetic phase transition (low-spin state to high-spin state) near -5.0 GPa with lattice relaxation, which is caused by increased atomic separation and localized Fe-\textit{d} electrons. Our results suggest that the electronic structures of Fe$_2$\textit{X}Si (\textit{X}=Ti and V) are very sensitive to external nonthermal parameters, which can be used to fine-tune physical properties for certain applications. Therefore, potential band engineering with chemical doping is suggested for future studies. 

\section{Acknowledgments}
H. K., C. K., and J. C. acknowledge the financial support from the STEM Center at Bergen Community College. N. H. and K. H. acknowledge the use of Expanse and Bridges at the San Diego Supercomputer Center (SDSC) and the Pittsburgh Supercomputing Center (PSC) through allocation TG-PHY190050 from the Advanced Cyberinfrastructure Coordination Ecosystem: Services and Support (ACCESS).

\bibliographystyle{apsrev4-2}
\bibliography{mybib_iso4}
\end{document}